\documentclass[aps,prc,reprint,floatfix,amsmath,amssymb]{revtex4-2}

\usepackage{graphicx}
\usepackage{bm}
\usepackage{xcolor}
\usepackage{amsmath}
\usepackage{amssymb}
\usepackage[colorlinks=true, breaklinks=true, linkcolor=blue, citecolor=purple, urlcolor=teal]{hyperref}

\renewcommand{\Im}{\mathop{\mathrm{Im}}}

\def\nuc#1#2{\relax\ifmmode{}^{#1}{\protect\text{#2}}\else${}^{#1}$#2\fi}

\begin{document}

\title{Self-consistent spectral framework for inclusive nonelastic breakup: The Trojan-horse method as its sub-Coulomb resonant limit}

\author{Jin Lei}
\email[]{jinl@tongji.edu.cn}
\affiliation{School of Physics Science and Engineering, Tongji University, Shanghai 200092, China}
\affiliation{Southern Center for Nuclear-Science Theory (SCNT), Institute of Modern Physics, Chinese Academy of Sciences, Huizhou 516000, Guangdong Province, China}

\date{\today}

\begin{abstract}
At the keV-scale energies of stellar nucleosynthesis, the resonant charged-particle reactions addressed by the Trojan-horse method (THM) proceed through isolated near-threshold resonances, so the low-energy cross section and the resonance strength carry the same information. The standard THM working formula is used there as a fixed extraction prescription, without a controlled reduction from the inclusive nonelastic breakup parent that would identify what it sets aside. I derive the sub-Coulomb resonant THM extraction directly from the Ichimura-Austern-Vincent inclusive nonelastic breakup cross section. A diagonal isolated-pole spectral ansatz for the absorptive participant-target potential, with four explicit validity conditions, reduces the inclusive cross section in the isolated-resonance limit to a per-pole distorted-wave Born approximation cross section on the resonance state, retaining full distortions and the postform interaction and weighted by the channel branching ratio; this is the operational quantity for resonance-strength extraction. The same one-level total-width profile follows independently from the postform transfer amplitude through the exact final states, so the extraction is route independent; a three-layer width analysis fixes its absorptive strength as half the nonelastic decay width, resolving the width and sign ambiguities of the literature. The standard factorized THM formula then follows under three dynamical approximations, plane waves, on-shell binary amplitude, and remnant neglect, together with a zero-range simplification of the projectile vertex, making explicit what it sets aside: the partial-wave coherence and the postform remnant.
\end{abstract}

\maketitle

%=====================================================
\section{Introduction}
\label{sec:intro}
%=====================================================

The Trojan-horse method (THM) is a principal indirect method for accessing low-energy charged-particle reactions at energies relevant to stellar nucleosynthesis, where the entrance channel is suppressed by the Coulomb barrier of the binary subreaction \cite{Baur1986,TypelBaur2003,NACRE1999,Tribble2014,Tumino2021,SolarFusionIII2025}. The present work addresses its resonant branch, in which the low-energy cross section is dominated by isolated near-threshold resonances and the quantity extracted is the resonance strength; the nonresonant energy-dependence applications that the method also serves are outside the present scope. For the keV-scale resonances considered here the energy dependence of the cross section is fixed by the resonance parameters \cite{LaneThomas1958}. The resonance strength and that energy dependence therefore carry the same information, and the resonance description is the theoretical content of that energy dependence, not an alternative to it.

The method extracts the strength of a binary subreaction $A(x,c)B$ from a quasifree three-body breakup $A(a,c\,b)B$ with $a=b+x$, in which the Trojan-horse $a$ delivers the participant $x$ to the target above the projectile Coulomb barrier while the spectator $b$ carries off the excess energy, so the $x+A$ subreaction occurs at sub-Coulomb relative energy. Its standard working formula is a plane-wave impulse approximation (PWIA) reduction of an underlying transfer matrix element \cite{LaCognata2010,LaCognata2015,Indelicato2017,Guardo2024,Petruse2025,Su2025}, with both distorted waves replaced by plane waves and the projectile structure entering through a Fourier transform.

The PWIA reduction is performed at the quasifree (QF) kinematic condition of THM measurements, a narrow cut on the spectator-participant momentum $\bm{p}_{bx}$ about zero, reconstructed event by event from the detected spectator; the explicit reconstruction and the distinction between this kinematic center and the computed cross-section maximum are collected in Appendix~\ref{app:kinematics}. The resulting factorized expression carries a half-off-energy-shell (HOES) binary amplitude whose continuation to the on-shell cross section is a separate step in the THM literature. The per-pole distorted-wave Born approximation (DWBA) cross section is instead evaluated with full entrance and exit distortions, so the quasifree factorization onto the projectile momentum distribution is recovered as the reduction rather than assumed.

The validity of the PWIA reduction at sub-Coulomb energies has been raised but not assessed in a controlled framework. The earliest off-shell treatment in the THM context is restricted to elastic two-body subreactions and s-wave separable potentials \cite{TuminoMukhamedzhanov2008}. A subsequent treatment extends the off-shell binary amplitude through a generalized $R$-matrix and surface-integral approach \cite{MukhamedzhanovKadyrov2010,Mukhamedzhanov2011,MukhamedzhanovKadyrovPang2020}, applied to the \nuc{19}{F}$\,+\,p$ system itself in Ref.~\cite{LaCognataMukhamedzhanov2011}. Neither treatment includes the entrance and exit distortions of the three-body process or the postform remnant of a non-Hermitian participant-target interaction. Reference~\cite{BHT2018} labeled the inclusive-to-PWIA reduction as a chain of approximations awaiting assessment, and the recent review~\cite{Tumino2025PPNP} presents the Ichimura-Austern-Vincent (IAV) inclusive cross section in closed form but does not project it onto the narrow-resonance per-pole limit or document the reduction chain to the factorized formula; that link has not appeared in the published THM literature.

The quantitative stakes have grown sharply for the \nuc{19}{F}$(p,\alpha\gamma)$\nuc{16}{O} reaction \cite{NACRE1999,Lombardo2015,ZhangRmatrix2021,deBoer2021,Guardo2024,Petruse2025,Redigolo2025,SolarFusionIII2025}, where a recent THM measurement of the 11 keV resonance strength \cite{Su2025} disagrees with the $R$-matrix evaluation by about a factor of six in the partial decay strength. The $R$-matrix value propagates JUNA direct and partner-channel data to the near-threshold pole through a global fit \cite{ZhangJUNA2021PRL,ZhangJUNA2022PRC,ZhangJUNA2022Nature,ChenJUNA2024,LiuJUNA2025Review}, and the discrepancy carries consequences for Population III calcium production \cite{deBoer2021,ZhangJUNA2022Nature}. This tension concentrates the validity question into a single benchmark.

The two frameworks differ in kind. The THM working formulas in standard use, plane-wave (PWIA) and modified distorted-wave Born approximation (MDWBA) alike, are extraction prescriptions: a measured coincidence yield in, a binary resonance strength out, the fixed reduction taken as given, with no direct sub-Coulomb cross-section data to test it. The IAV inclusive nonelastic breakup framework adopted here is instead a predictive calculation, the reaction dynamics in and the per-pole cross section out, comparable to measurement and already tested above the Coulomb barrier \cite{LeiMoro2015,PotelNunesThompson2015,CarlsonCapoteSin2016,LiuLeiRen2023}. The present work derives the extraction prescription as the approximate reduction of this predictive cross section, making the discarded content explicit and assessable.

The observable also differs in kind. The inclusive parent is differential only in the spectator kinematics $(\Omega_b,E_b)$ and integrates the $x+A$ decay angle, so it delivers the angle-integrated resonance strength directly and removes the off-shell angular continuation of the binary amplitude that the factorized reduction must perform. The spectator angular distribution it retains instead carries sensitivity to the resonance spin-parity $J_n^\pi$, at the cost of the binary decay angular correlation.

The IAV framework gives the inclusive nonelastic breakup cross section of a three-body breakup with a complex participant-target optical potential as a Green's-function expectation value of its imaginary part \cite{IAV1985,HusseinMcVoy1985,AusternCDCC1987}, and has since been developed with controlled post-prior equivalence, an angular-momentum-basis extension, and continuum-discretized wave functions \cite{LeiMoro2015Reexamining,LeiMoro2015,LeiMoro2018,Lei2018AngularMomentum,LeiMoro2019,LeiMoro2023}. Reference~\cite{LeiMoro2018} derives a Breit-Wigner spectral decomposition, a sum over narrow poles of the participant-target Green's function $G_x(E_x)=(E_x-T_x-U_x)^{-1}$ at $E_x=E_n-i\xi_n^\text{IAV}$ in the quasibound limit (the absorptive half-width, made precise in Sec.~\ref{sec:width-dict}), each identified one to one with a resonance state $\phi_n$; in the narrow-resonance limit the inclusive cross section reduces to a Lorentzian-weighted single-pole transfer matrix element at the resonance energy, the natural starting point for sub-Coulomb resonant THM extraction.

Existing numerical work stops short of the sub-Coulomb narrow-pole benchmark. Reference~\cite{MukhamedzhanovPangKadyrov2019} formulates the THM cross section through a surface-integral off-shell decomposition but in production evaluates only a spectator-weighted zero-range DWBA ratio in which the off-shell structure cancels, so its conclusion on the sub-Coulomb $S$-factor rests on the ratio rather than the full inclusive parent. The IAV-revival groups \cite{PotelNunesThompson2015,CarlsonCapoteSin2016,Potel2017,CarlsonFredericoHussein2017} implement the IAV inclusive formalism at intermediate and high energies, none targeting a sub-Coulomb THM benchmark or the narrow-resonance projection.

The present work builds the analytical framework connecting the parent IAV inclusive cross section to the THM extraction at sub-Coulomb energies. A diagonal isolated-pole spectral ansatz on the absorptive participant-target potential carries four explicit validity conditions, each reducible to a checkable estimate: two are closed dimensionless bounds from $R$-matrix tabulations, one is a model-dependent continuum-decoupling diagnostic, and one is a normalization diagnostic on the pole basis. Under this ansatz the inclusive cross section reduces, in the isolated-resonance limit, to a single-pole Breit-Wigner contribution that recovers the THM resonance-strength extraction with the partial-wave coherence and postform remnant retained. A three-layer width analysis combining Feshbach projection with $R$-matrix normalization fixes the width relations and resolves the corresponding width ambiguities of the literature. The framework is self-consistent in the sense that a single absorptive operator $W_x$ both defines the inclusive yield and, in the sub-Coulomb limit, sets the Lorentzian width of each pole, so the extraction formula is derived from the parent cross section rather than postulated alongside it. The same spectral form is then rederived directly from the postform transfer amplitude, through the one-level resonance profile of the exact participant-target final states, with no reference to the inclusive closed form; the two routes yield the same one-level total-width profile, so the per-pole extraction is route independent.

The paper is organized as follows. Section~\ref{sec:iav-recap} recapitulates the IAV framework and its Breit-Wigner spectral form; Sec.~\ref{sec:ansatz} introduces the diagonal isolated-pole spectral ansatz and its validity conditions; Sec.~\ref{sec:width-dict} derives the width relations and channel projection; Sec.~\ref{sec:direct} rederives the spectral form directly from the postform transfer amplitude. Section~\ref{sec:regimes} establishes the isolated-resonance limit and the per-pole extraction recipe; Sec.~\ref{sec:postform} gives the postform remnant decomposition and the reduction to the factorized PWIA-THM formula; Sec.~\ref{sec:summary} summarizes.

%=====================================================
\section{IAV inclusive nonelastic breakup and its Breit-Wigner spectral form}
\label{sec:iav-recap}
%=====================================================

I consider the three-body process
\begin{equation}
a + A \;\to\; b + (x+A)^{*}, \qquad a = b+x,
\label{eq:reaction}
\end{equation}
an inclusive nonelastic breakup reaction in which the projectile $a=b+x$ separates into a detected spectator $b$ and an unobserved residual system $B^{*}\equiv(x+A)^{*}$ that denotes any (bound, resonant, or continuum) nonelastic final state of the $x+A$ composite. The participant $x$ undergoes a nonelastic interaction with the target $A$, and only the spectator $b$ is observed. The projectile is incident at an energy above the $a+A$ Coulomb barrier, and this above-barrier entrance is the \emph{only} kinematic requirement of the present framework. The subsystem relative energy $E_x$ is then fixed by energy conservation from the detected spectator kinematics $(\Omega_b,E_b)$. The spectator energies that map to small $E_x$ reach the sub-Coulomb, even sub-threshold, region of the $x+A$ subsystem, and because the participant $x$ is delivered to the target by the above-barrier projectile rather than tunneling into the $x+A$ channel, that region is reached without the entrance Gamow suppression of a direct measurement. The quasifree condition $p_{bx}\!\approx\!0$ is not imposed in the general per-pole description of this Trojan-horse mechanism; it re-enters when the same description is reduced to the standard PWIA-THM extraction in Sec.~\ref{sec:postform}. The inclusive nonelastic breakup cross section, summed over all final states of the $x+A$ subsystem except the elastic $x+A$ channel, was derived in Ref.~\cite{IAV1985} in post form and reads
\begin{equation}
\frac{d^2\sigma_\text{NEB}}{d\Omega_b\, dE_b}
= -\frac{2}{\hbar v_a}\,\rho_b(E_b)\,\langle\psi_x(\bm{k}_b)|\,W_x\,|\psi_x(\bm{k}_b)\rangle,
\label{eq:iav-inclusive}
\end{equation}
where $v_a=\hbar k_a/\mu_a$ is the projectile velocity in the entrance channel with relative momentum $k_a$ and reduced mass $\mu_a$, $\rho_b(E_b)=\mu_b k_b/[(2\pi)^3\hbar^2]$ is the spectator phase-space density at spectator relative momentum $k_b$ and reduced mass $\mu_b$ in the exit channel, $W_x=\Im U_x$ is the imaginary part of the complex participant-target optical potential $U_x=V_x^R+iW_x$, and $\psi_x(\bm{k}_b,\bm{r}_x)$ is the participant wave function defined as the solution of an inhomogeneous Schr\"odinger equation with $U_x$ and outgoing-wave boundary conditions, with the source set by the postform interaction \cite{IAV1985,LeiMoro2015,LeiMoro2018}. The asymptotic outgoing-wave amplitude of $\psi_x$ in the $x+A$ elastic channel carries the elastic-breakup (EBU) content of the three-body process, in which $A$ remains in its ground state, while the matrix element $\langle\psi_x(\bm{k}_b)|W_x|\psi_x(\bm{k}_b)\rangle$ projects out the flux removed from this elastic channel into the nonelastic decay channels, so that Eq.~(\ref{eq:iav-inclusive}) counts only the nonelastic breakup yield. I adopt the postform throughout; Eq.~(\ref{eq:iav-inclusive}) is the canonical postform inclusive cross section and the starting point of the present analysis. It is exact within the effective three-body model at the DWBA-level source, that is, within the participant-target optical potential $U_x$, the entrance and exit distorted waves, and the projectile bound state. Those inputs are evaluated at well-separated kinematics where smooth parametrizations are standard, so the only genuinely sub-Coulomb-specific assumption is the structure of $W_x$ on the resonance scale, addressed below. The matrix element $\langle\psi_x|W_x|\psi_x\rangle$ carries the entire $x+A$ dynamical content; it is what all subsequent approximations act on.

In the narrow-resonance regime the spectral structure of the participant-target propagator is dominated by a discrete set of isolated resonance states $\{\phi_n\}$ in the energy window. The symbol $\phi_n$ is used throughout for the single-particle resonance form factor in the standard Lane-Thomas $R$-matrix interior convention~\cite{LaneThomas1958,Tribble2014}: the interior eigenfunction of $T_x+V_x^R$ satisfying $(T_x+V_x^R)\phi_n=E_n\phi_n$ on the region $r\le a$ with the Bloch boundary operator imposed at $r=a$, where $V_x^R$ is a real participant-target potential adjusted so that the eigenvalue equals the experimental resonance energy $E_n$, and $u_l(r)=r\phi_n(r)$ is normalized to unit probability over $r\le a$. This unit normalization is the single-particle ($C^2S=1$) convention adopted throughout the main development; the general spectroscopic-factor rescaling $\phi_n=\sqrt{C^2S}\,\hat\phi_n$ is reinstated in Sec.~\ref{sec:regimes} where the formation cross section is built. I adopt the observed-parameter convention: the Bloch boundary condition is fixed pole by pole so that the real $R$-matrix level shift vanishes at the resonance energy, and the tabulated observed widths and level energies are used directly as inputs. The residue factor relating formal to observed parameters, and the explicit diagnostic that keeps it close to unity, are given in Sec.~\ref{sec:direct} and Appendix~\ref{app:bounds}.

The form factors $\{\phi_n\}$ are therefore not $L^2$ eigenstates of the full participant-target Hamiltonian above the $x+A$ threshold, and they are not mutually orthogonal in the $L^2$ sense across the channel surface; they are interior-normalized form factors on which the per-pole DWBA transfer matrix element acts. Beyond $r=a$, $\phi_n$ is continued by its asymptotic Coulomb tail, matched at $r=a$ to the surface amplitude $u_l(a)$, through which the standard $R$-matrix relation~(\ref{eq:gamma-p-sp}) fixes the partial widths.

Inserting the spectral expansion of the participant-target propagator on the pole set $\{\phi_n\}$ into Eq.~(\ref{eq:iav-inclusive}) and reorganizing, Lei and Moro~\cite{LeiMoro2018} derived the spectral form rigorously for constant $W_x$ in the quasibound limit, the pole displaced by the absorption alone; the same step extends to diagonal $W_x$ on the pole basis, with perturbative corrections from the off-diagonal couplings, and keeping the elastic escape in the pole displacement (Appendix~\ref{app:width}) refines the denominator to the total half-width. The diagonality is promoted to a controlled ansatz in Sec.~\ref{sec:ansatz}. The spectral form reads
\begin{equation}
\frac{d^2\sigma_\text{NEB}}{d\Omega_b\, dE_b}
= \sum_n \omega_n(E_x)\,\frac{d\sigma_n^\text{DWBA}}{d\Omega_b},
\label{eq:iav-spectral}
\end{equation}
where $E_x$ is the participant-target subsystem energy reconstructed from the measured spectator energy $E_b$, and the spectral weight $\omega_n(E_x)$ is the Lorentzian profile
\begin{equation}
\omega_n(E_x) = \frac{\xi_n^\text{IAV}/\pi}{(E_x-E_n)^2+(\Gamma_n^\text{tot}/2)^2},
\label{eq:omega-n}
\end{equation}
peaked at $E_x=E_n$ with full width at half-maximum $\Gamma_n^\text{tot}$ and absorptive numerator $\xi_n^\text{IAV}$, both made precise in Sec.~\ref{sec:width-dict}. The quasibound form of Ref.~\cite{LeiMoro2018}, with $\xi_n^\text{IAV}$ alone in the denominator, is recovered from Eq.~(\ref{eq:omega-n}) in the sub-Coulomb limit $\Gamma_x^n\to 0$; the same total-width profile is rederived independently from the full participant-target final states in Sec.~\ref{sec:direct}. The pole transfer cross section appearing in Eq.~(\ref{eq:iav-spectral}) is built from the standard distorted-wave Born approximation transfer matrix element evaluated on the resonance state $\phi_n$,
\begin{align}
\frac{d\sigma_n^\text{DWBA}}{d\Omega_b}
&= \frac{2\pi\,\rho_b}{\hbar v_a}\,\left|\mathcal{M}_n^\text{DWBA}\right|^2, \nonumber \\
\mathcal{M}_n^\text{DWBA}
&\equiv \langle \phi_n\,\chi_b^{(-)}|\,V_\text{post}\,|\,\chi_a^{(+)}\,\phi_a\rangle,
\label{eq:Mn-DWBA}
\end{align}
in which $\chi_a^{(+)}(\bm{k}_a,\bm{r}_a)$ is the entrance distorted wave generated by the projectile-target optical potential, $\chi_b^{(-)}(\bm{k}_b,\bm{r}_b)$ is the exit distorted wave generated by the spectator-residual optical potential, $\phi_a$ is the projectile internal wave function in the bound configuration $a=b+x$, and $V_\text{post}=V_{bx}+(U_{bA}-U_{bB})$ is the postform interaction (here $V_{bx}$ is the spectator-participant interaction inside the projectile, $U_{bA}$ is the spectator-target optical potential, and $U_{bB}$ is the spectator-residual optical potential). The matrix element $\mathcal{M}_n^\text{DWBA}$ of Eq.~(\ref{eq:Mn-DWBA}) is a volume integral. Its participant-target radial integral, over which the resonance form factor $\phi_n$ carries the asymptotic tail of the $x+A$ subsystem, is evaluated by the Vincent-Fortune complex-contour method~\cite{VincentFortune1970,Lei2025PRC}, which defines $\mathcal{M}_n^\text{DWBA}$ unambiguously with no auxiliary surface construction. For the $E_n>0$ resonances the external $\phi_n$ is the outgoing Coulomb function $H_l^{+}(\eta_n,k_n r_x)$ at real wave number $k_n$, the width entering only through the Lorentzian denominator of Eq.~(\ref{eq:omega-n}) and not as a complex wave number in $\phi_n$; the contour is taken in the damping sector of the asymptotic Coulomb product. All per-pole DWBA cross sections in this paper are summed over the final magnetic substates and averaged over the $(2j_a+1)(2j_A+1)$ initial substates set by the projectile and target ground-state spins $j_a$ and $j_A$; the explicit spin-projection structure is suppressed in Eq.~(\ref{eq:Mn-DWBA}) and reinstated in Eq.~(\ref{eq:kappa-def}). The properties of $V_\text{post}$, and why the prior form is not under controlled approximation at sub-Coulomb energies, are taken up in Sec.~\ref{sec:postform}.

A resonance $n$ can decay through several open channels, including elastic $x+A$ re-emission with participant-channel partial width $\Gamma_x^n$, and the nonelastic decay channels (e.g.~$\alpha$ and $\gamma$ for the \nuc{19}{F}$\,+\,p$ resonances) with partial widths $\Gamma_c^n$ for $c\ne x$, and total width $\Gamma_n^\text{tot}=\Gamma_x^n+\Gamma_n^\text{non-el}$ where $\Gamma_n^\text{non-el}=\sum_{c\ne x}\Gamma_c^n$. The absorptive operator $W_x$ removes flux from the elastic channel into the nonelastic decay channels of the resonance; the absorptive strength $\xi_n^\text{IAV}$ in the numerator of Eq.~(\ref{eq:omega-n}) is therefore the diagonal matrix element of $W_x$ on $\phi_n$, $\xi_n^\text{IAV}=|\langle\phi_n|W_x|\phi_n\rangle|$. Its further identification with the sum of nonelastic partial widths, $\xi_n^\text{IAV}=\Gamma_n^\text{non-el}/2$, is the Lane-Thomas $R$-matrix result derived below. Equation~(\ref{eq:Mn-DWBA}) makes this separation explicit at the amplitude level: the common breakup source $\langle\chi_b^{(-)}|V_\text{post}|\chi_a^{(+)}\phi_a\rangle$ projected onto the $x+A$ elastic scattering state gives the elastic-breakup amplitude, while its projection onto $\phi_n$ is the nonelastic counterpart counted by $\langle\psi_x|W_x|\psi_x\rangle$. The form factor $\phi_n$ is thus the resonance pole through which the reaction proceeds, and $\Gamma_x^n$ stays on the elastic-breakup side, absent from $W_x$ and from $\xi_n^\text{IAV}$.

Equations~(\ref{eq:iav-spectral})-(\ref{eq:Mn-DWBA}) identify, per resonance, the dynamical content of the inclusive cross section: a Lorentzian profile set by the participant-target propagator alone, and a coherent partial-wave transfer matrix element set by the entrance and exit distortions and by the resonance state $\phi_n$. By partial-wave coherence I mean that $\mathcal{M}_n^\text{DWBA}$ is a phase-coherent sum over the orbital angular momenta that link the entrance and exit distorted waves to the $(l,j)$ components of the resonance form factor $\phi_n$, with the relative phases of the terms retained; the factorized reduction of Sec.~\ref{sec:postform} collapses this coherent three-body coupling structure onto a single plane-wave, on-shell term, whereas the per-pole amplitude keeps it in full.

%=====================================================
\section{Diagonal isolated-pole spectral ansatz of the absorptive optical potential}
\label{sec:ansatz}
%=====================================================

The spectral form~(\ref{eq:iav-spectral}) is exact only when the absorptive participant-target potential acts diagonally on the chosen resonance-state pole basis and is decoupled from the genuine continuum of the participant-target subsystem. In the sub-Coulomb astrophysical regime the absorptive potential is not smooth on the resonance scale: it is dominated by sparse narrow resonances of the participant-target subsystem rather than by the smooth megaelectronvolt-scale optical average that justifies smooth-$W_x$ optical models at higher energies. I therefore introduce the spectral form as an explicit ansatz with four validity conditions that I make testable rather than assume.

The ansatz is the diagonal isolated-pole representation of the absorptive kernel restricted to the pole subspace,
\begin{equation}
W_x \;\simeq\; \sum_n |\phi_n\rangle\,W_n\,\langle\phi_n|,
\label{eq:wx-ansatz}
\end{equation}
in which $\{\phi_n\}$ is the set of isolated resonance states of the participant-target subsystem in the energy window and $W_n=\langle\phi_n|W_x|\phi_n\rangle$ is the diagonal absorptive matrix element on the corresponding pole form factor. The matrix element $W_n$ encodes the absorption of the resonance state into the nonelastic decay channels, that is, into the complement of the elastic $x+A$ channel that $W_x$ removes flux from. Its identification with the nonelastic decay-width sum $\Gamma_n^\text{non-el}/2$, the quantity that sets the absorptive numerator of the inclusive yield, is established in Sec.~\ref{sec:width-dict} through a three-layer width analysis.

Equation~(\ref{eq:wx-ansatz}) is therefore not a global spectral representation of the optical potential at all energies but the diagonal pole truncation of $W_x$ within the resonance subspace spanned by $\{|\phi_n\rangle\}$ in the narrow energy interval where the isolated resonances dominate (the rank-$N$ pole projection); the off-diagonal and form-factor non-orthogonality corrections are bounded in Appendix~\ref{app:bounds}. The ansatz is controlled to the stated accuracy when the following four conditions hold:
\par\smallskip\noindent\textbf{Condition 1. Diagonality of $W_x$ on the pole basis.} The off-diagonal matrix elements $W_{nm}=\langle\phi_n|W_x|\phi_m\rangle$ for $n\neq m$ vanish in the exact isolated-pole limit; the controlled practical requirement is $|W_{nm}|/|E_n-E_m|\ll 1$.
\par\smallskip\noindent\textbf{Condition 2. Continuum decoupling.} In addition to the resonance states $\{\phi_n\}$, the participant-target subsystem has a smooth, nonresonant continuum of $x+A$ scattering states $\{\phi_E\}$ that fills the energy gaps between the resonances. Condition 2 asserts that all matrix elements of $W_x$ that touch a $\phi_E$ on either side, $\langle\phi_n|W_x|\phi_E\rangle$ or $\langle\phi_E|W_x|\phi_{E'}\rangle$, are negligible against the diagonal pole elements $W_n=\langle\phi_n|W_x|\phi_n\rangle$. Physically, the absorption flows entirely through the discrete resonances, so the pole expansion~(\ref{eq:wx-ansatz}) misses no absorptive strength.
\par\smallskip\noindent\textbf{Condition 3. Pole isolation.} Each resonance $n$ has a total half-width $\Gamma_n^\text{tot}/2$, equal at sub-Coulomb energies to the absorptive $\xi_n^\text{IAV}=|W_n|$ (Sec.~\ref{sec:width-dict}), small against the spacing to the nearest neighbor.
\par\smallskip\noindent\textbf{Condition 4. Normalization of the pole basis.} The interior-normalized form factors carry unit pole residues and negligible mutual overlaps, $Z_n\simeq 1$ and $\max_{n\neq m}|S_{nm}|\ll 1$ [Eq.~(\ref{eq:norm-diagnostic})], with $Z_n$ the formal-to-observed residue defined in Sec.~\ref{sec:direct}; the working formulas of Secs.~\ref{sec:width-dict}--\ref{sec:postform} hold to relative accuracy $O(1-Z_n)$ and $O(\max|S_{nm}|)$.
\par\smallskip
Conditions 1 and 3 are governed by two related but distinct dimensionless ratios, the Cauchy-Schwarz off-diagonal bound $\sqrt{\xi_n^\text{IAV}\xi_m^\text{IAV}}/|E_n-E_m|$ of Eq.~(\ref{eq:smallness-parameter}) and the isolation ratio $(\Gamma_n^\text{tot}/2)/|E_n-E_m|$; both close from experimental partial widths and level energies alone. Condition 2 is instead a model-dependent diagnostic, the continuum-to-resonance reaction-cross-section ratio $\epsilon_{\rm cont}$, which must be assessed within a specified continuum model rather than asserted from Coulomb penetration, since the entrance penetrability cancels in the ratio. Condition 4 is likewise not controlled by the width-to-spacing ratio, since $\Gamma_c^n=2P_c\gamma_{nc}^2$ can be penetration suppressed while $\gamma_{nc}^2|S_c'|$ is not, and closes from tabulated reduced widths and Coulomb shift functions. The supporting bounds [Eqs.~(\ref{eq:schwarz}), (\ref{eq:smallness-parameter}), (\ref{eq:norm-diagnostic}), and~(\ref{eq:epsilon-cont})] are derived in Appendix~\ref{app:bounds}.

When the four conditions hold to their stated accuracy, the diagonal action of $W_x$ on the pole basis $\{|\phi_n\rangle\}$ converts the matrix element $\langle\psi_x(\bm{k}_b)|W_x|\psi_x(\bm{k}_b)\rangle$ in Eq.~(\ref{eq:iav-inclusive}) into the spectral sum~(\ref{eq:iav-spectral}) with the Lorentzian profile~(\ref{eq:omega-n}). The spectral form is therefore a narrow-resonance approximation, with the condition-2 model dependence absorbed into the systematic uncertainty of the ansatz. Its narrow-resonance limit, in which $\omega_n$ collapses to the branching-weighted delta of Sec.~\ref{sec:regimes}, recovers the THM extraction and is the focus of the present analysis. For light-nucleus narrow-resonance reactions of astrophysical interest the widths and level energies needed for the condition-1 and condition-3 ratios are supplied directly by comprehensive $R$-matrix evaluations~\cite{NACRE1999,deBoer2021,SolarFusionIII2025}.

%=====================================================
\section{Width relations, integrated-pole-area, and channel projection}
\label{sec:width-dict}
%=====================================================

The literature on indirect methods for narrow-resonance extraction suffers from three recurring ambiguities: between the half-width $\Gamma/2$ and the full width $\Gamma$ at half-maximum, between sign conventions for the imaginary part of the optical potential, and between the absorptive half-width $\xi_n^\text{IAV}$ and the experimental partial decay widths. The third is the substantive one: $\xi_n^\text{IAV}$ is defined through Eqs.~(\ref{eq:iav-spectral})-(\ref{eq:omega-n}), whereas $\Gamma_x^n$ and $\Gamma_c^n$ govern the branching of the resonance into its participant-channel and nonelastic decay channels. I make these correspondences explicit before deriving any quantitative result.

The convention adopted throughout is that the imaginary part of the absorptive optical potential is non-positive, $W_x\leq 0$, so that the operator $-W_x$ is positive semidefinite and absorbs flux from the elastic channel. The diagonal matrix element $\langle\phi_n|W_x|\phi_n\rangle\equiv W_n$ is therefore non-positive, and I define the absorptive half-width as the positive quantity $\xi_n^\text{IAV}\equiv|W_n|=-W_n>0$; the partial widths $\Gamma_x^n$, $\Gamma_c^n$ and the total width $\Gamma_n^\text{tot}$ follow the uppercase full-width-at-half-maximum (FWHM) $\Gamma$ convention. The original derivation of Ref.~\cite{LeiMoro2018} worked with the signed $W_n$ directly; the correspondence between the two conventions is the single substitution $W_n=-\xi_n^\text{IAV}$.

In the narrow-resonance limit, the absorptive half-width is one half the nonelastic partial-width sum, which coincides with one half the experimental total decay width at sub-Coulomb energies where $\Gamma_x^n\ll\Gamma_n^\text{non-el}$,
\begin{equation}
\xi_n^\text{IAV} \;=\; \tfrac{1}{2}\,\Gamma_n^\text{non-el}
\;=\; \tfrac{1}{2}\sum_{c\ne x}\Gamma_c^n
\;\simeq\; \tfrac{1}{2}\,\Gamma_n^\text{tot},
\label{eq:width-dict}
\end{equation}
in which $\Gamma_n^\text{tot}=\Gamma_x^n+\Gamma_n^\text{non-el}$ is the experimental total decay width of the resonance $n$ (for the \nuc{19}{F}$\,+\,p$ application, $x=p$ and the nonelastic channels are $\alpha$ and $\gamma$). The strict identity of Eq.~(\ref{eq:width-dict}) is to the nonelastic sum alone, since $\Gamma_x^n$ is the elastic re-emission carried by the elastic-breakup branch and therefore absent from the absorptive $W_x$; the approximate equality to $\tfrac{1}{2}\Gamma_n^\text{tot}$ holds because $\Gamma_x^n$ is Coulomb-suppressed at sub-Coulomb energies for charged $x$, with residual relative correction $\Gamma_x^n/\Gamma_n^\text{tot}$. $\Gamma_x^n$ re-enters the formalism as the formation coupling of the resonance strength [Eq.~(\ref{eq:omega-gamma-def}), Layer C below], not as part of $\xi_n^\text{IAV}$; the loose phrasing ``$W_x$ absorbs into all open channels'' conflates the two roles~\cite{LeiMoro2018,Tribble2014}. Equation~(\ref{eq:width-dict}) is established by a three-layer width analysis combining Feshbach projection with $R$-matrix normalization. Layer A places the pole of the IAV Green's function $G_x=(E_x-T_x-U_x)^{-1}$ on $\phi_n$ at $E_x=E_n-i\Gamma_n^\text{tot}/2$, the sum of the elastic escape half-width and the absorptive displacement $\xi_n^\text{IAV}=|\langle\phi_n|W_x|\phi_n\rangle|$. Layer B identifies the same matrix element, through the imaginary part of the Lane-Thomas level matrix, with the nonelastic partial-width sum $\Gamma_n^\text{non-el}/2$. Layer C fixes the overall normalization of $\phi_n$ entering the per-pole amplitude of Eq.~(\ref{eq:Mn-DWBA}) through the participant-channel entrance reduced width $\gamma_x^2$. Layers A and B are derived in Appendix~\ref{app:width}. A residual elastic-channel correction of order $\Gamma_x^n/\Gamma_n^\text{tot}$ arises only if the quasibound substitution $\xi_n^\text{IAV}\simeq\Gamma_n^\text{tot}/2$ is used in place of the full profile: negligible at the $11$~keV resonance, at the percent level at the higher reference resonances, where any use of the substitution must be propagated in the systematic budget (Appendix~\ref{app:width}). Layer C carries the only dynamical content beyond the widths and is given here.

\emph{Layer C: Formation amplitude through the entrance reduced width.} The per-pole DWBA amplitude of Eq.~(\ref{eq:Mn-DWBA}) is linear in the overall normalization of $\phi_n$, controlled by the entrance reduced width $\gamma_x^2$, the third distinct width of the formalism, not by the absorptive matrix element of $W_x$. For a single participant-channel single-particle configuration ($C^2S=1$) with $\phi_n$ the interior eigenfunction at the experimental $E_n$, the surface amplitude $u_l(a)$ is fixed by the standard Lane-Thomas single-particle reduced-width relation
\begin{equation}
\Gamma_x^{\,\text{sp}} \;=\; 2\,P_l(E_n,a)\,\gamma_x^{\,2}, \qquad \gamma_x^{\,2} \;=\; \frac{\hbar^2}{2\mu_{xA}\,a}\,|u_l(a)|^2,
\label{eq:gamma-p-sp}
\end{equation}
with $P_l(E_n,a)$ the standard $R$-matrix penetration factor at orbital angular momentum $l$~\cite{LaneThomas1958,Tribble2014}, $a$ the participant-channel radius, $\mu_{xA}$ the participant-target reduced mass, and $u_l(r)=r\,\phi_n(r)$ the unit-normalized interior form factor. The observed partial width is $\Gamma_x^n=C^2S\,\Gamma_x^{\,\text{sp}}$: the single-particle potential model does not by itself reproduce the tabulated width, and $C^2S$ carries the mismatch. The first pass sets $C^2S=1$, the convention of the analysis of Ref.~\cite{Su2025}, so that the unit-probability normalization of $u_l$ fixes both $\gamma_x^2$ and the overall normalization of $\phi_n$ entering Eq.~(\ref{eq:Mn-DWBA}); the extraction of Sec.~\ref{sec:regimes} is independent of this choice, since $C^2S$ cancels in the kernel $\kappa_n$ [Eq.~(\ref{eq:kappa-def})]. The per-pole cross section therefore scales as $|\mathcal{M}_n^\text{DWBA}|^2\propto\gamma_x^2\propto\Gamma_x^n$ at fixed $E_n$ [Eq.~(\ref{eq:gamma-p-sp})], modulo the distorted-wave matching of $\chi_a^{(+)},\chi_b^{(-)}$ to $\phi_n$. The formation amplitude is thus controlled by $\Gamma_x^n$, not by the absorptive strength $\xi_n^\text{IAV}$, which sets the nonelastic numerator and already aggregates the nonelastic channels [Eqs.~(\ref{eq:width-dict}),~(\ref{eq:Wx-nonelastic})]; conflating the two would double count. This proportionality holds only in the single-particle limit; beyond it $\phi_n$ carries coherent partial-wave admixture that $\gamma_x^2$ alone does not encode but the per-pole amplitude retains in full~\cite{Lei2018AngularMomentum}.

The Lorentzian profile~(\ref{eq:omega-n}) integrates over the full real-energy axis to the nonelastic branching fraction,
\begin{equation}
\int_{-\infty}^{+\infty} dE_x\,\omega_n(E_x) = \frac{\Gamma_n^\text{non-el}}{\Gamma_n^\text{tot}}\;\simeq\;1,
\label{eq:integrated-area}
\end{equation}
the last approximation in the sub-Coulomb regime. The energy-integrated formulation makes the resonance-strength extraction independent of the half-width versus full-width convention. In an actual measurement the spectator gate, threshold cuts, and finite resolution restrict the integration to a finite window and convolve it with an instrumental response, an experimental-side modification outside the present scope~\cite{Tribble2014,Tumino2025PPNP}.

The experimental observable in a typical resonant THM application is the exclusive channel-gated yield of a specific decay channel $c$ of the participant-target subsystem. The projection from the inclusive nonelastic breakup of Eq.~(\ref{eq:iav-spectral}) onto this exclusive channel follows the formation-decay separation of an isolated resonance. With $\sigma_n^\text{form}$ the per-pole DWBA cross section of Eq.~(\ref{eq:Mn-DWBA}) integrated over the spectator angle $\Omega_b$, the energy integral~(\ref{eq:integrated-area}) retains the fraction of the formed resonance that decays nonelastically, $\sigma_n^\text{NEB}=(\Gamma_n^\text{non-el}/\Gamma_n^\text{tot})\,\sigma_n^\text{form}$, and the channel gate selects $\Gamma_c^n/\Gamma_n^\text{non-el}$ of that, composing to
\begin{equation}
\sigma_n^{\text{excl},c} \;=\; \frac{\Gamma_c^n}{\Gamma_n^\text{non-el}}\,\sigma_n^\text{NEB} \;=\; \frac{\Gamma_c^n}{\Gamma_n^\text{tot}}\,\sigma_n^\text{form} \;\equiv\; b_c^n\,\sigma_n^\text{form},
\label{eq:exclusive-projection}
\end{equation}
with the standard formation-decay branching ratio $b_c^n=\Gamma_c^n/\Gamma_n^\text{tot}$ defined relative to the experimental total width. In the sub-Coulomb regime where $\Gamma_n^\text{non-el}\simeq\Gamma_n^\text{tot}$, the intermediate factor collapses to unity and the experimentally familiar form $\sigma_n^{\text{excl},c}\simeq b_c^n\,\sigma_n^\text{NEB}$ is recovered. The factorization~(\ref{eq:exclusive-projection}), taken in the energy-integrated sense, is exact within the scalar isolated-pole branching approximation; corrections enter through the off-diagonal absorptive couplings $W_{nm}$ and through level interference, and are of the same order as the Cauchy-Schwarz bound~(\ref{eq:schwarz}) on $W_{nm}$.

%=====================================================
\section{Direct derivation of the spectral form from the postform transfer amplitude}
\label{sec:direct}
%=====================================================

The spectral form~(\ref{eq:iav-spectral}) was reached above by expanding the participant-target propagator inside the inclusive closed form~(\ref{eq:iav-inclusive}). The same per-pole structure follows from a general postform DWBA expression alone, with no reference to Eq.~(\ref{eq:iav-inclusive}) or to the absorptive operator $W_x$, by singling out the resonance profile of the exact participant-target final states. This section gives that derivation; the width correspondence of Sec.~\ref{sec:width-dict} is invoked afterwards only to compare the two results, not within the derivation itself. The two routes give the same one-level total-width profile: the spectral form is the narrow-resonance structure of the postform transfer amplitude itself, not an artifact of either construction.

Consider the breakup gated on one nonelastic decay channel $c$ of the participant-target subsystem. The postform DWBA amplitude for populating the channel-$c$ scattering state $\Psi_{c,E_x}^{(-)}$ of the full participant-target Hamiltonian at subsystem energy $E_x$ is $\langle\Psi_{c,E_x}^{(-)}\chi_b^{(-)}|V_\text{post}|\chi_a^{(+)}\phi_a\rangle$, built from the same entrance and exit distorted waves $\chi_a^{(+)},\chi_b^{(-)}$, projectile bound state $\phi_a$, and postform interaction $V_\text{post}$ as Eq.~(\ref{eq:Mn-DWBA}), with $\Psi_{c,E_x}^{(-)}$ normalized per unit energy. The channel-gated double-differential cross section, summed over the unobserved quantum numbers of channel $c$, reads
\begin{equation}
\frac{d^2\sigma^{\text{excl},c}}{d\Omega_b\,dE_b}
=\frac{2\pi\rho_b}{\hbar v_a}\,\big|\langle\Psi_{c,E_x}^{(-)}\chi_b^{(-)}|V_\text{post}|\chi_a^{(+)}\phi_a\rangle\big|^2,
\label{eq:direct-exclusive}
\end{equation}
with $E_x$ fixed by the detected spectator kinematics as throughout.

Near an isolated narrow resonance the channel-$c$ scattering state admits the one-level decomposition of $R$-matrix theory~\cite{LaneThomas1958,Tribble2014}, in the same observed-parameter convention as Sec.~\ref{sec:iav-recap}, valid over the full volume: $\Psi_{c,E_x}^{(-)}=A_{c,n}(E_x)\,\tilde\phi_n+\Psi^\text{bg}$, where $\tilde\phi_n$ is the resonance state, the interior eigenfunction continued at $r=a$ by its matched Coulomb tail, $A_{c,n}$ is the energy-dependent level amplitude, and $\Psi^\text{bg}$ is the smooth nonresonant background, neglected here. The amplitude then factorizes as
\begin{equation}
\langle\Psi_{c,E_x}^{(-)}\chi_b^{(-)}|V_\text{post}|\chi_a^{(+)}\phi_a\rangle
\simeq\langle\Psi_{c,E_x}^{(-)}|\phi_n\rangle_a\,\mathcal{M}_n^\text{DWBA},
\label{eq:direct-factorization}
\end{equation}
with $\mathcal{M}_n^\text{DWBA}$ the full-volume per-pole matrix element of Eq.~(\ref{eq:Mn-DWBA}), evaluated by the same Vincent-Fortune contour, and $\langle\Psi_{c,E_x}^{(-)}|\phi_n\rangle_a=A_{c,n}^{*}(E_x)$ its coefficient. The subscript denotes the $R$-matrix internal product over $r\le a$, taken in the $x+A$ channel component of $\Psi_{c,E_x}^{(-)}$; it measures the level amplitude because $\tilde\phi_n$ restricted to $r\le a$ is the unit-normalized $\phi_n$ ($\phi_n$ is not an $L^2$ state, so an unrestricted overlap would not converge). No localization of the transfer source is invoked: the factorization rests on the one-level dominance of $\Psi_{c,E_x}^{(-)}$, not on confinement of the source to the interior. The projection amplitude carries the entire resonant energy dependence and obeys the standard one-level spectral-strength profile
\begin{equation}
\big|\langle\phi_n|\Psi_{c,E_x}^{(-)}\rangle_a\big|^2
=\frac{\Gamma_c^n/2\pi}{(E_x-E_n)^2+(\Gamma_n^\text{tot}/2)^2},
\label{eq:spectral-strength}
\end{equation}
whose energy integral is the branching ratio $\Gamma_c^n/\Gamma_n^\text{tot}=b_c^n$ of Eq.~(\ref{eq:exclusive-projection}) and whose sum over all open channels integrates to unity: the unit spectral strength of the quasi-stationary state $\phi_n$ is distributed over its open decay channels in proportion to the partial widths~\cite{LaneThomas1958}. The channel projection of Sec.~\ref{sec:width-dict} is thus reproduced here without a separate formation-decay argument.

The unit total strength asserted in Eq.~(\ref{eq:spectral-strength}) is the residue of the one-level pole on the interior-normalized $\phi_n$, and it is unity only in the observed-parameter convention. For a formal $R$-matrix level the residue is instead $Z_n\equiv\big(1+\sum_c\gamma_{nc}^{2}\,S_c'(E_n)\big)^{-1}$, dimensionless because the reduced widths $\gamma_{nc}^2$ carry energy and the shift-function derivatives $S_c'$ its inverse; this is the standard factor converting formal to observed parameters~\cite{LaneThomas1958,Tribble2014}. The $\gamma_{nc}^2$ entering $Z_n$ are the formal reduced widths; evaluating them from the tabulated observed widths through $\Gamma_c^n=2P_c\gamma_{nc}^2$ is itself accurate to the same order $1-Z_n$ that the diagnostic monitors, so the diagnostic is self-consistent at its own accuracy. Adopting observed parameters throughout, as fixed in Sec.~\ref{sec:iav-recap}, refers Eq.~(\ref{eq:spectral-strength}) and the surface relation~(\ref{eq:gamma-p-sp}) to one and the same normalization of $\phi_n$, neglecting corrections of order $1-Z_n$. The proximity of $Z_n$ to unity is an explicit diagnostic, not a consequence of narrowness: $\Gamma_c^n=2P_c\gamma_{nc}^2$ can be small through the penetrability alone while $\gamma_{nc}^2|S_c'|$ remains finite, this requirement is condition 4 of Sec.~\ref{sec:ansatz}, closed from tabulated reduced widths and Coulomb shift functions in Appendix~\ref{app:bounds}. Inserting Eqs.~(\ref{eq:direct-factorization}) and~(\ref{eq:spectral-strength}) into Eq.~(\ref{eq:direct-exclusive}) and summing over the nonelastic channels $c\neq x$ and over the isolated poles of the window gives
\begin{align}
\frac{d^2\sigma_\text{NEB}}{d\Omega_b\,dE_b}
&=\sum_n\tilde\omega_n(E_x)\,\frac{d\sigma_n^\text{DWBA}}{d\Omega_b},\nonumber\\
\tilde\omega_n(E_x)&=\frac{\Gamma_n^\text{non-el}/2\pi}{(E_x-E_n)^2+(\Gamma_n^\text{tot}/2)^2}.
\label{eq:direct-spectral}
\end{align}

Equation~(\ref{eq:direct-spectral}) coincides with the spectral form~(\ref{eq:iav-spectral})-(\ref{eq:omega-n}): the numerators identically, $\Gamma_n^\text{non-el}/2\pi=\xi_n^\text{IAV}/\pi$ by Eq.~(\ref{eq:width-dict}), and the denominators through the same total half-width, which the absorptive route acquires from the elastic escape in the pole displacement (Appendix~\ref{app:width}). The ratio $\Gamma_x^n/\Gamma_n^\text{tot}$ does not separate the two routes; it measures only the quasibound approximation of Ref.~\cite{LeiMoro2018}, in which the escape is dropped from the denominator: $\approx 2\times10^{-31}$ at the $11$~keV resonance of \nuc{19}{F}$\,+\,p$, and at the percent level at the higher reference resonances (Appendix~\ref{app:width}).

The assumptions consumed by the direct route are two: one-level dominance of the scattering state, and neglect of the smooth nonresonant background. These parallel the validity conditions of Sec.~\ref{sec:ansatz}: pole isolation (condition 3) substantiates the one-level step and continuum decoupling (condition 2) the background neglect; condition 1, the diagonality of $W_x$, is specific to the absorptive representation and has no counterpart here. The per-pole extraction formula of Sec.~\ref{sec:regimes} therefore does not rest on the inclusive closed form~(\ref{eq:iav-inclusive}): whatever reservations one may hold about an absorptive optical-model representation at sub-Coulomb energies, the same extraction quantity emerges from the exact final states directly. The inclusive parent remains the natural starting point, fixing what the spectral sum excludes, the elastic-breakup branch; the absorptive route in addition supplies, through Eq.~(\ref{eq:width-dict}), the operational reading of the absorptive strength from tabulated partial widths.

%=====================================================
\section{Isolated-resonance limit, the per-pole extraction formula, and outlook on the overlapping-resonance regime}
\label{sec:regimes}
%=====================================================

The Lorentzian spectral form~(\ref{eq:iav-spectral})-(\ref{eq:omega-n}) admits two regime limits that are distinguished by the ratio $(\Gamma_n^\text{tot}/2)/|E_n-E_m|$ of the total half-width to the nearest-neighbor level spacing. The isolated-resonance limit, $\Gamma_n^\text{tot}/2\ll|E_n-E_m|$, is the regime relevant to sub-Coulomb THM extraction and is the focus of this section; the overlapping-resonance regime, $\Gamma_n^\text{tot}/2\gtrsim|E_n-E_m|$, lies outside the scope of the diagonal isolated-pole ansatz and is discussed at the end of the section as outlook.

In the isolated-resonance limit, taken for the nearest neighbor of any spin and parity, the Lorentzian profile~(\ref{eq:omega-n}) of vanishing width collapses in the distributional sense to a Dirac delta on the resonance energy weighted by the nonelastic branching,
\begin{equation}
\omega_n(E_x)\;\xrightarrow[\Gamma_n^\text{tot}\to 0]{}\;\frac{\Gamma_n^\text{non-el}}{\Gamma_n^\text{tot}}\,\delta(E_x-E_n),
\label{eq:delta-collapse}
\end{equation}
which combined with the channel gate $\Gamma_c^n/\Gamma_n^\text{non-el}$ reproduces the composition of Eq.~(\ref{eq:exclusive-projection}) exactly. Operationally, integrated against any test function smooth on the level-spacing scale over a window $\Gamma_n^\text{tot}/2\ll\Delta E\ll|E_n-E_m|$, the spectral weight acts as $\int_{\Delta E}dE_x\,\omega_n(E_x)\,F(E_x)\simeq (\Gamma_n^\text{non-el}/\Gamma_n^\text{tot})\,F(E_n)$. The spectral sum~(\ref{eq:iav-spectral}) then becomes a sum over single-pole transfer matrix elements evaluated at $E_x=E_n$.

Combining the isolated-resonance limit~(\ref{eq:delta-collapse}) and the channel projection~(\ref{eq:exclusive-projection}), the per-pole exclusive spectator yield in the spectator window takes the form
\begin{equation}
\frac{d^2\sigma_n^{\text{excl},c}}{d\Omega_b\, dE_b}
\;=\; b_c^n \,\,\frac{d\sigma_n^\text{DWBA}}{d\Omega_b}\;\delta(E_x-E_n),
\label{eq:thm-recovery}
\end{equation}
in which $d\sigma_n^\text{DWBA}/d\Omega_b$ of Eq.~(\ref{eq:Mn-DWBA}) is evaluated on $\phi_n$ with full entrance and exit distortions and the postform interaction, and $b_c^n$ projects the inclusive yield onto the gated decay channel $c$; the accuracy of the delta collapse~(\ref{eq:delta-collapse}) is controlled by conditions 1--3, its overall normalization by condition 4.

The quantity an actual extraction evaluates is not the delta-valued density of Eq.~(\ref{eq:thm-recovery}) itself but its energy integral over the spectator window. Integrating over the detected spectator energy $E_b$ across the resonance, and writing $\delta(E_x-E_n)=\delta(E_b-E_b^\ast)\,|dE_x/dE_b|^{-1}$ with $E_b^\ast$ the spectator energy that reconstructs $E_x=E_n$ by energy conservation from the detected spectator kinematics, collapses the delta and leaves a finite single-differential cross section,
\begin{equation}
\frac{d\sigma_n^{\text{excl},c}}{d\Omega_b}
=\int_{\Delta E} dE_b\,\frac{d^2\sigma_n^{\text{excl},c}}{d\Omega_b\,dE_b}
= b_c^n\,\frac{d\sigma_n^\text{DWBA}}{d\Omega_b}\,\left|\frac{dE_x}{dE_b}\right|^{-1}_{E_b^\ast}.
\label{eq:thm-integrated}
\end{equation}
The integration is over the spectator energy $E_b$ (equivalently the relative energy $E_x$); the level energy $E_n$ is fixed and is not integrated. In the relative variables of the main text the Jacobian is not an approximation: energy conservation fixes $E_x$ linearly by $E_b$ at fixed spectator angle, so $|dE_x/dE_b|=1$ identically, and the factor is kept in Eq.~(\ref{eq:thm-integrated}) only to mark where the laboratory-frame Jacobian of Appendix~\ref{app:kinematics} enters when the yield is compared with detected spectator energies. The finite single-differential cross section on the left of Eq.~(\ref{eq:thm-integrated}) is the quantity the method delivers; the delta in Eq.~(\ref{eq:thm-recovery}) only fixes its energy location at $E_x=E_n$, equivalently the spectator energy $E_b^\ast$ at which $d\sigma_n^\text{DWBA}/d\Omega_b$ is evaluated.

The astrophysical resonance strength of a decay channel $c$ is defined, following the conventions of Refs.~\cite{NACRE1999,Tribble2014,SolarFusionIII2025}, as
\begin{equation}
(\omega\gamma)_n^c \;\equiv\; \frac{2J_n+1}{(2j_x+1)(2j_A+1)}\,\frac{\Gamma_x^n\,\Gamma_c^n}{\Gamma_n^\text{tot}},
\label{eq:omega-gamma-def}
\end{equation}
with $J_n$ the total angular momentum of the resonance $n$ in the compound $x+A$ system (uppercase) and $j_x,j_A$ the participant and target ground-state spins (lowercase, on the same convention as the entrance-channel spins $j_a,j_A$). The statistical factor $(2J_n+1)/[(2j_x+1)(2j_A+1)]$ is the spin weight for forming this resonance in the binary $x+A\to n$ subreaction. The compound symbol $\omega\gamma$ is the standard astrophysical resonance-strength notation, not a product of the spectral Lorentzian $\omega_n(E_x)$ of Eq.~(\ref{eq:omega-n}) with a width $\gamma$.

The differential cross section of Eq.~(\ref{eq:thm-integrated}) integrates over the spectator angle to the channel yield $\sigma_n^{\text{excl},c}=b_c^n\,\sigma_n^\text{form}$ of Eq.~(\ref{eq:exclusive-projection}), and this yield connects to the strength of Eq.~(\ref{eq:omega-gamma-def}) through the formation cross section.

In the single-particle first-pass the per-pole DWBA cross section is linear in $\Gamma_x^n$ through the form-factor normalization established in Layer C [Eq.~(\ref{eq:gamma-p-sp})], so the angle-integrated formation cross section is $\sigma_n^\text{form}=\kappa_n\,\Gamma_x^n$, with $\kappa_n$ the per-pole DWBA formation cross section per unit participant width. Writing $\phi_n=\sqrt{C^2S}\,\hat\phi_n$ with $\hat\phi_n$ the unit-normalized single-particle form factor ($\int_0^a|\hat u_l|^2\,dr=1$, $\hat u_l=r\hat\phi_n$), so that the observed partial width is $\Gamma_x^n=C^2S\,\Gamma_x^\text{sp}$ with $\Gamma_x^\text{sp}$ the single-particle width of Eq.~(\ref{eq:gamma-p-sp}) built on $\hat\phi_n$, and $\widehat{\mathcal{M}}_n=\langle\hat\phi_n\,\chi_b^{(-)}|V_\text{post}|\chi_a^{(+)}\,\phi_a\rangle$ the matrix element built on it, the kernel is explicitly
\begin{equation}
\kappa_n=\frac{1}{\Gamma_x^\text{sp}}\,\frac{2\pi\rho_b}{\hbar v_a}\int d\Omega_b\,\overline{\big|\widehat{\mathcal{M}}_n(\Omega_b)\big|^2},
\label{eq:kappa-def}
\end{equation}
where the overbar denotes the average over the $(2j_a+1)(2j_A+1)$ entrance substates and the sum over the final substates. The matrix element $\widehat{\mathcal{M}}_n$ and the single-particle width $\Gamma_x^\text{sp}$ are built from the same $\hat\phi_n$, so the $C^2S$ normalization cancels in the ratio and $\kappa_n$ carries only the distorted-wave matching, the spectator phase space, and the entrance-channel spin average, independent of the overall form-factor normalization. Inserting $\sigma_n^\text{form}=\kappa_n\,\Gamma_x^n$ into the branching relation gives
\begin{equation}
\sigma_n^{\text{excl},c}=\kappa_n\,\frac{\Gamma_x^n\,\Gamma_c^n}{\Gamma_n^\text{tot}}=\kappa_n\,\frac{(2j_x+1)(2j_A+1)}{2J_n+1}\,(\omega\gamma)_n^c.
\label{eq:strength-bridge}
\end{equation}
Equation~(\ref{eq:strength-bridge}) is the explicit bridge between the three-body reaction the method runs and the binary reaction it constrains: the left side is the three-body channel yield the IAV-DWBA computation delivers, the right side the binary $x+A$ strength built from widths and spins alone, and $\kappa_n$ carries the entire three-body content. The two spin averages do not cancel: $\kappa_n$ averages over the projectile entrance channel $a+A$, weight $(2j_a+1)(2j_A+1)$, whereas $(\omega\gamma)_n^c$ carries the participant formation weight $(2j_x+1)(2j_A+1)$, and $j_a\neq j_x$ in general. Equation~(\ref{eq:strength-bridge}) runs in three modes. Directly, a single resonance suffices: an absolutely normalized yield divided by the computed $\kappa_n$ and the spin factor returns $(\omega\gamma)_n^c$, with no reference resonance. Relative to a reference resonance of known strength, factors common to the levels, such as the absolute detection efficiency, cancel, and only $(2J_n+1)$ and $\kappa_n$ vary; this is the standard THM procedure used by the measurement motivating this work~\cite{Su2025}, and it remains preferable whenever the absolute scale is the dominant systematic. The direct mode is an alternative extraction permitted by Eq.~(\ref{eq:strength-bridge}); it does not alter the role of reference normalization in the standard THM procedure. In forward-consistency mode all widths are fixed from tabulations and the predicted yield tests the framework instead of extracting anything. Equations~(\ref{eq:thm-integrated})-(\ref{eq:strength-bridge}), evaluated with the width inputs of Sec.~\ref{sec:width-dict}, constitute the computational recipe of the per-pole extraction, valid to the relative accuracy $O(1-Z_n)$ and $O(\max|S_{nm}|)$ of condition 4.

In the complementary regime $\Gamma_n^\text{tot}/2\gtrsim|E_n-E_m|$, relevant to compound-nucleus reactions at higher level density, the diagonal ansatz~(\ref{eq:wx-ansatz}) no longer holds and the spectral sum cannot be evaluated pole by pole. Under energy averaging, random off-diagonal phases, and a local-density approximation it would instead reorganize into the Hauser-Feshbach surrogate-method form~\cite{HauserFeshbach1952,Escher2012RMP}; its controlled derivation is left to future work.

%=====================================================
\section{Controlled postform source and reduction to the factorized PWIA-THM formula}
\label{sec:postform}
%=====================================================

\begin{figure}[!t]
\centering
\includegraphics[width=\columnwidth]{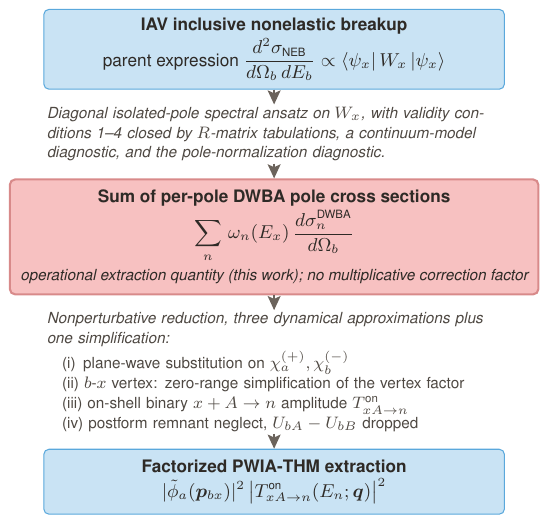}
\caption{Two-step nonperturbative reduction from the IAV inclusive parent [Eq.~(\ref{eq:iav-inclusive})] to the factorized PWIA-THM formula [Eq.~(\ref{eq:pwa-reduction})]~\cite{LaCognata2010,Tribble2014,Su2025}. The per-pole DWBA cross section at the intermediate level (highlighted) is the operational extraction quantity; the factorized expression is a further reduction of the same parent, not a multiplicative correction. The spectral ansatz and conditions 1--4 are introduced in Sec.~\ref{sec:ansatz}; the reduction steps, three dynamical approximations plus a zero-range vertex simplification, are detailed here.}
\label{fig:reduction-chain}
\end{figure}

I now turn to the postform structure of the per-pole amplitude and its reduction to the factorized PWIA-THM expression. The transfer matrix element of Eq.~(\ref{eq:Mn-DWBA}) is the starting point of the second reduction in Fig.~\ref{fig:reduction-chain}, which lists the reduction steps to the factorized expression; the prior form is uncontrolled at sub-Coulomb energies in its standard smooth-potential version, for the two independent reasons developed below. The channel branching ratio $b_c^n$ and the Lorentzian denominator are identical in the per-pole amplitude and in this reduction, so they cancel in the operational comparison, and the entire difference between $(\omega\gamma)_n^\text{DWBA}$ and $(\omega\gamma)_n^\text{PWIA-THM}$, the strength of Eq.~(\ref{eq:omega-gamma-def}) evaluated through the two forms, is carried by the steps below, of which (i), (iii), and (iv) hold the dynamical content. The postform interaction is
\begin{equation}
V_\text{post}\;=\;V_{bx}\;+\;(U_{bA}-U_{bB}),
\label{eq:Vpost}
\end{equation}
in which the projectile binding interaction $V_{bx}$ is the direct breakup term and the difference $U_{bA}-U_{bB}$ is the postform remnant, which arises because the spectator-target distortion in the entrance arrangement differs from the spectator-residual distortion in the exit arrangement.

The postform expression is adopted not by preference but because the prior form is uncontrolled at sub-Coulomb energies on two counts. First, completeness: for a complex participant-target potential the post and prior per-pole amplitudes differ by the nonorthogonality (Hussein-McVoy) amplitude $\Delta\mathcal{M}_n\equiv\mathcal{M}_n^\text{post}-\mathcal{M}_n^\text{prior}$~\cite{HusseinMcVoy1985,IAV1985,LeiMoro2015,LeiMoro2018}, the per-pole cross-section difference being
\begin{equation}
|\mathcal{M}_n^\text{post}|^2-|\mathcal{M}_n^\text{prior}|^2
=2\,\mathrm{Re}\!\big(\mathcal{M}_n^{\text{prior}\,*}\,\Delta\mathcal{M}_n\big)+|\Delta\mathcal{M}_n|^2,
\label{eq:postprior-square}
\end{equation}
whose cross term survives after squaring: $\phi_n$ diagonalizes only the real $V_x^R$ while the channel retains the full $U_x$, whose absorptive part carries the resonance's own nonelastic decay, so the bare prior reduction is incomplete unless $\Delta\mathcal{M}_n$ and the cross term are restored. Second, constructability: the prior operator $V_\text{prior}=U_x+(U_{bA}-U_{aA})$ carries that full complex $U_x$, whose absorptive part is the sparse isolated-resonance spectrum available only through the spectral ansatz of Eq.~(\ref{eq:wx-ansatz}), never a smooth $W_x(r)$ that a global parametrization~\cite{KoningDelaroche2003} could supply; a controlled prior would demand a pole-by-pole assembly of $U_x$, outside the present scope. The postform operator~(\ref{eq:Vpost}) carries no participant-target interaction at all, the resonance entering only through $\phi_n$; built from $V_{bx}$ and the smooth spectator potentials, it is complete and constructible where the prior form is neither.

The first plane-wave limit of the transfer matrix element of Eq.~(\ref{eq:Mn-DWBA}) is obtained by replacing the entrance and exit distorted waves by plane waves,
\begin{equation}
\chi_a^{(+)}(\bm{k}_a,\bm{r}_a)\to e^{i\bm{k}_a\cdot\bm{r}_a},\qquad
\chi_b^{(-)}(\bm{k}_b,\bm{r}_b)\to e^{-i\bm{k}_b\cdot\bm{r}_b}.
\label{eq:pwa-substitution}
\end{equation}
This combined plane-wave substitution defines a plane-wave Born transfer cross section,
\begin{align}
\frac{d\sigma_n^\text{PWBA}}{d\Omega_b}&=
\frac{2\pi\,\rho_b}{\hbar v_a}\,\big|\mathcal{M}_n^\text{PWBA}\big|^2,\nonumber\\
\mathcal{M}_n^\text{PWBA}&=\langle \phi_n\,e^{-i\bm{k}_b\cdot\bm{r}_b}|\,V_\text{post}\,|\,e^{i\bm{k}_a\cdot\bm{r}_a}\,\phi_a\rangle.
\label{eq:Mn-PWBA}
\end{align}
Equation~(\ref{eq:Mn-PWBA}) is not yet the factorized PWIA-THM expression used in resonance-strength extraction. Beyond the plane-wave substitution (i) that defines it, three further reductions are required. (ii) The $b$-$x$ vertex factor is reduced to its zero-range constant; by the bound-state identity given below the finite-range vertex already factorizes as a smooth kinematic factor times $\tilde\phi_a$ for any local $V_{bx}$, so this step is a simplification of that factor, not a structural approximation. (iii) The binary subreaction amplitude is evaluated on shell, at the kinematic point set by the spectator-window cuts. (iv) Only the $V_{bx}$ piece of $V_\text{post}$ is retained; the remnant $U_{bA}-U_{bB}$ is dropped, justified when the two spectator distortions are similar, $U_{bA}\simeq U_{bB}$. Carrying out these reductions makes the factorization explicit. In the projectile Jacobi coordinates $(\bm{r}_{bx},\bm{r}_{xA})$ the entrance and exit plane-wave phases separate into a spectator-participant part and a participant-target part, and retaining only the $V_{bx}$ piece [step (iv)] the amplitude of Eq.~(\ref{eq:Mn-PWBA}) factorizes as
\begin{align}
\mathcal{M}_n^\text{PWBA}\big|_{V_{bx}}
&=\Big[\int d\bm{r}_{bx}\,e^{-i\bm{p}_{bx}\cdot\bm{r}_{bx}}\,V_{bx}(\bm{r}_{bx})\,\phi_a(\bm{r}_{bx})\Big]\nonumber\\
&\quad\times\Big[\int d\bm{r}_{xA}\,\phi_n^{*}(\bm{r}_{xA})\,e^{i\bm{q}\cdot\bm{r}_{xA}}\Big],
\label{eq:pwa-factorization}
\end{align}
with spectator-participant momentum $\bm{p}_{bx}=\bm{k}_b-(m_b/m_a)\bm{k}_a$ and momentum transfer $\bm{q}=\bm{k}_a-\bm{k}_b$. The participant-target phase carries $\bm{q}$ in the no-recoil (heavy-core) limit: the exact conjugate momentum of $\bm{r}_{xA}$ is $\bm{k}_a-\frac{m_A}{m_x+m_A}\bm{k}_b$, which reduces to $\bm{q}$ as $m_x/m_A\to 0$. The remnant neglect of step (iv) is customarily argued in the same heavy-core limit, where $A$ and $B=A+x$ generate similar spectator distortions, but $U_{bA}-U_{bB}$ depends on the charges and potential geometries as well as on the mass ratio, so the two corrections are related, not controlled by a single parameter. The two brackets have a direct physical reading. The first is the $b$-$x$ breakup vertex: by the bound-state Schr\"odinger equation it equals $-(\hbar^2/2\mu_{bx})(p_{bx}^2+\kappa_a^2)\,\tilde\phi_a(\bm{p}_{bx})$, with $\kappa_a$ the bound-state wave number of $\phi_a$, so it is the momentum-space wave function $\tilde\phi_a$ times a smooth kinematic factor, constant for a zero-range $V_{bx}$ [step (ii)] and collected into $\mathcal{N}_\text{kin}$ below. The second is the momentum-space overlap of the resonance form factor $\phi_n$, the formation vertex of the binary $x+A\to n$ subreaction; step (iii) is the statement that this vertex, evaluated at the on-shell point set by the spectator-window cuts, is identified with the on-shell binary $T$-matrix $T_{xA\to n}^\text{on}(E_n;\bm{q})$, an identification that defines the approximation rather than following from the factorization. Squaring this factorized amplitude in $d\sigma_n^\text{PWBA}/d\Omega_b$ and collecting the prefactor $2\pi\rho_b/\hbar v_a$ together with the $b$-$x$ breakup-vertex constant into $\mathcal{N}_\text{kin}$ gives the standard factorized form
\begin{align}
\frac{d\sigma_n^\text{PWIA-THM}}{d\Omega_b}
&=\mathcal{N}_\text{kin}(\Omega_b,k_b)\,
|\tilde\phi_a(\bm{p}_{bx})|^2 \nonumber\\
&\quad\times
\left|T_{xA\to n}^\text{on}(E_n;\bm{q})\right|^2,
\label{eq:pwa-reduction}
\end{align}
in which $\mathcal{N}_\text{kin}$ collects the spectator-window phase-space, spin-average, and frame factors, and the remaining symbols are as defined above. At the resonance peak $|T_{xA\to n}^\text{on}|^2$ carries the resonance formation strength $\propto\Gamma_x^n$ of the participant channel; the decay branching to the detected channel $c$ enters once, through the channel projection $b_c^n$ of Eqs.~(\ref{eq:exclusive-projection}) and~(\ref{eq:thm-recovery}), so that $b_c^n$ and $|T_{xA\to n}^\text{on}|^2$ together assemble the astrophysical resonance strength $(\omega\gamma)_n^c$ of Eq.~(\ref{eq:omega-gamma-def}). 

The factorization~(\ref{eq:pwa-reduction}) holds algebraically at any $\bm{p}_{bx}$ once steps (i)-(iv) are taken; the quasifree cut $\bm{p}_{bx}\approx 0$ selects the region where $|\tilde\phi_a(\bm{p}_{bx})|^2$ is largest and the impulse reading of the spectator is cleanest, a kinematic event selection rather than a dynamical approximation. In the standard treatment the quasifree spectator picture is the premise from which the factorization is argued; here the factorized form follows from the four operator-level approximations, and the quasifree condition re-enters only as the kinematic point at which the form is evaluated. The dynamical content therefore sits in the reduction steps, not in the selection. The clearest case is the remnant $U_{bA}-U_{bB}$ dropped at step (iv), the standard postform remnant of transfer and breakup DWBA: it is suppressed for heavy targets where $U_{bA}\simeq U_{bB}$, but non-negligible for a light target such as \nuc{19}{F}, and it is a more consequential departure from the per-pole amplitude than the quasifree selection that the standard analyses place at the center.

Equations~(\ref{eq:Mn-PWBA}) and~(\ref{eq:pwa-reduction}) expose the factorized PWIA-THM working formula of Refs.~\cite{LaCognata2010,Su2025} as a \emph{nonperturbative reduction} of the per-pole DWBA amplitude: four finite steps, not a perturbative limit interpolating between the two, and no multiplicative correction factor relating the two forms. Steps (i), (iii), and (iv) each remove specific dynamical content, the entrance and exit distortions, the off-shell structure of the binary vertex, and the remnant; step (ii) only simplifies the smooth vertex factor.

The intermediate plane-wave Born cross section of Eq.~(\ref{eq:Mn-PWBA}) is a stage of this reduction, not a baseline against which the per-pole cross section is a multiplicative distortion correction. The plane-wave substitution strips three convolved layers from $\chi_a^{(+)},\chi_b^{(-)}$ at once, Coulomb wave, surface phase shift, and interior amplitude, so no graded ordering reconstructs the distorted amplitude from the plane-wave one; a $(d,pX)$ radial-cut-off study finds the interior alone worth $20$--$25\%$ even above the barrier~\cite{LiuLeiRen2023}. Nor does the sub-Coulomb suppression factor out: the universal Gamow penetrability enters the on-shell $|T_{xA\to n}^\text{on}|^2$ of Eq.~(\ref{eq:pwa-reduction}) through the standard $R$-matrix $P_l$, whereas the distortion residue is set by the matching of $\chi_a^{(+)},\chi_b^{(-)}$ to $\phi_n$, whose small-radius structure is nonperturbative at $\eta_{xA}\gg 1$.

%=====================================================
\section{Summary and outlook}
\label{sec:summary}
%=====================================================

I have established four results for the IAV inclusive nonelastic breakup cross section in the sub-Coulomb resonant regime.

First, the diagonal isolated-pole spectral ansatz of the absorptive participant-target optical potential is controlled in the narrow-resonance regime once the condition-1 and condition-3 ratios from $R$-matrix tabulations, the model-dependent continuum diagnostic $\epsilon_{\rm cont}$, and the condition-4 normalization diagnostic $Z_n\simeq 1$, $\max_{n\neq m}|S_{nm}|\ll 1$ [Eq.~(\ref{eq:norm-diagnostic})] are evaluated.

Second, a three-layer width analysis combining Feshbach projection with $R$-matrix normalization establishes the width relations $\xi_n^\text{IAV}=\Gamma_n^\text{non-el}/2\simeq\Gamma_n^\text{tot}/2$, resolving the half-width-versus-full-width and sign ambiguities of the literature and the recurring conflation of the absorptive half-width with the participant-channel partial width.

Third, the inclusive cross section reduces in the isolated-resonance limit to a per-pole distorted-wave Born approximation transfer matrix element evaluated on the resonance state with full entrance and exit distortions and the postform interaction, weighted by the channel branching ratio. The postform source is the central convention at sub-Coulomb energies because the prior-form alternative would require an explicit pole-by-pole assembly of $U_x$ outside the scope of the present formalism.

Fourth, the same one-level total-width profile follows directly from the postform transfer amplitude through the exact participant-target final states, with no reference to the inclusive closed form; the two routes coincide, the ratio $\Gamma_x^n/\Gamma_n^\text{tot}$ measuring only the quasibound approximation, and the per-pole extraction is route independent.

The central operational statement follows from these results. The standard factorized THM working formula and the per-pole DWBA cross section are \emph{successive reductions of the same IAV inclusive nonelastic breakup parent}: the isolated-resonance limit reduces the parent to the per-pole cross section, and three dynamical approximations with a zero-range vertex simplification further reduce that to the factorized form. The operational quantity for sub-Coulomb resonance-strength extraction is the per-pole cross section itself, which retains the partial-wave coherence of the direct $V_{bx}$ matrix element and the remnant $U_{bA}-U_{bB}$ that the reduction suppresses. The extraction computes $b_c^n\,d\sigma_n^\text{DWBA}/d\Omega_b$ following the recipe of Sec.~\ref{sec:regimes}, with the observed $R$-matrix partial widths as the only resonance inputs, alongside the optical potentials, distorted waves, and projectile bound state common to any DWBA calculation; the absorptive strength $\xi_n^\text{IAV}$ sets the nonelastic numerator and condition 1, the total half-width $\Gamma_n^\text{tot}/2$ the Lorentzian $E_x$ width and condition 3, and neither enters the integrated single-differential cross section beyond the branching ratio $b_c^n$. The offset of the resulting $(\omega\gamma)_n^\text{DWBA}$ from the factorized-THM value is set by the three dynamical approximations; its numerical evaluation at a specific benchmark, together with the closure of the condition-2 and normalization diagnostics, is left to the follow-up application.

%=====================================================
\begin{acknowledgments}
This work was supported by the National Natural Science Foundation of China (Grants No.~12475132 and No.~12535009) and the Fundamental Research Funds for the Central Universities. Large language model assistants were used for editorial polishing of the English text and for cross checking intermediate algebraic steps in the analytical derivations; every derivation and numerical statement was subsequently verified by hand by the author, who takes full responsibility for all content.
\end{acknowledgments}
%=====================================================

%=====================================================
\section*{Data availability}
%=====================================================

There are no publicly available research data or software supporting this paper. Requests for further information or data should be sent to the author.

\appendix

%=====================================================
\section{Quasifree kinematics: reconstruction of the spectator-participant momentum and the QF window}
\label{app:kinematics}
%=====================================================

This appendix reconstructs the quasifree (QF) window in standard nonrelativistic THM kinematics~\cite{Tumino2021}; laboratory quantities carry a ``$\text{lab}$'' label to distinguish them from the relative variables of the main text. The target $A$ is at rest, the beam $a=b+x$ (with $m_a=m_b+m_x$) carries laboratory momentum $\bm{k}_a^\text{lab}$ along $\hat{\bm z}$ with energy $E_\text{lab}=(k_a^\text{lab})^2/2m_a$, and the spectator $b$ is detected at laboratory angle $\theta_b^\text{lab}$ with kinetic energy $E_b^\text{lab}$, so $k_b^\text{lab}=\sqrt{2m_b E_b^\text{lab}}$ and $\bm{k}_b^\text{lab}\cdot\bm{k}_a^\text{lab}=k_b^\text{lab} k_a^\text{lab}\cos\theta_b^\text{lab}$.

In the spectator picture $b$ is emitted without interacting, so momentum conservation ($\bm{k}_b^\text{lab}+\bm{k}_x^\text{lab}=\bm{k}_a^\text{lab}$) splits the projectile laboratory momentum into the spectator center-of-mass share and the frame-independent internal Jacobi momentum, reconstructed event by event from the measured spectator momentum and the beam,
\begin{equation}
\bm{p}_{bx}=\bm{k}_b^\text{lab}-\frac{m_b}{m_a}\,\bm{k}_a^\text{lab},
\label{eq:app-pbx}
\end{equation}
with magnitude, in terms of the detected $(\theta_b^\text{lab},E_b^\text{lab})$,
\begin{equation}
p_{bx}^2=(k_b^\text{lab})^2-2\,\frac{m_b}{m_a}\,k_b^\text{lab} k_a^\text{lab}\cos\theta_b^\text{lab}+\Big(\frac{m_b}{m_a}\Big)^{\!2} (k_a^\text{lab})^2.
\label{eq:app-pbx-mag}
\end{equation}

The QF condition $\bm{p}_{bx}=0$ requires $\bm{k}_b^\text{lab}=(m_b/m_a)\bm{k}_a^\text{lab}$, the spectator coasting at the beam velocity. It fixes the QF point in both detected variables,
\begin{equation}
\theta_b^{\text{lab,QF}}=0,\qquad E_b^{\text{lab,QF}}=\frac{m_b}{m_a}\,E_\text{lab}.
\label{eq:app-qf}
\end{equation}
By Eq.~(\ref{eq:app-pbx-mag}) the QF window, the cut $p_{bx}\le p_\text{cut}$, is a finite region of the $(\theta_b^\text{lab},E_b^\text{lab})$ plane centered on Eq.~(\ref{eq:app-qf}), a near-forward angular acceptance together with an energy bite about $E_b^{\text{lab,QF}}$, with extent set by $p_\text{cut}$ and by the intrinsic width of $|\tilde\phi_a(\bm{p}_{bx})|^2$.

The point $\bm{p}_{bx}=0$ is the kinematic center of the window, where $|\tilde\phi_a|^2$ peaks; it is not imposed as the location of the cross-section maximum. Since the per-pole DWBA cross section is computed over the whole window with the distortions retained, the predicted yield maximum may be displaced from $\bm{p}_{bx}=0$, and that displacement is an output of the calculation measuring the departure from the impulse picture.

The binary relative energy $E_x$, the variable scanned for the excitation function, is reconstructed from the full three-body final state through the standard quasifree energy relation~\cite{Tumino2021,LaCognata2010}. The per-pole cross section is computed in the relative variables of the main text and converted to $(\theta_b^\text{lab},E_b^\text{lab})$ by the standard frame Jacobian when compared with data.

%=====================================================
\section{Validity bounds for the diagonal isolated-pole spectral ansatz}
\label{app:bounds}
%=====================================================

This appendix derives the dimensionless bounds for the conditions 1--4 of the spectral ansatz~(\ref{eq:wx-ansatz}).

I treat $W_x$ as a scalar (central) absorptive potential; non-central components such as imaginary spin-orbit are kept out as a working assumption, since global parametrizations~\cite{KoningDelaroche2003} average over precisely the narrow-resonance structure the ansatz resolves and cannot adjudicate their size there. With the absorptive convention $W_x\le 0$ (so that $-W_x$ is a positive-semidefinite quadratic form), the off-diagonal couplings $W_{nm}$ vanish for unequal spin-parity by rotational invariance, and for equal spin-parity satisfy the Cauchy-Schwarz bound
\begin{align}
 |W_{nm}|^2
&\leq \langle\phi_n|(-W_x)|\phi_n\rangle\,
        \langle\phi_m|(-W_x)|\phi_m\rangle \nonumber\\
&= \xi_n^\text{IAV}\,\xi_m^\text{IAV}.
\label{eq:schwarz}
\end{align}
The diagonal-ansatz controlling ratio is therefore
\begin{equation}
\frac{|W_{nm}|}{|E_n-E_m|} \;\leq\; \frac{\sqrt{\xi_n^\text{IAV}\,\xi_m^\text{IAV}}}{|E_n-E_m|},
\label{eq:smallness-parameter}
\end{equation}
small whenever the participant-target subsystem is in the narrow-resonance regime; the explicit evaluation at any specific reference resonance requires only the experimental partial widths and level energies of the participant-target subsystem.

The bound~(\ref{eq:schwarz}) controls the absorptive couplings but not the overlaps $S_{nm}=\langle\phi_n|\phi_m\rangle_a$ of the form factors themselves, which the pole expansion~(\ref{eq:wx-ansatz}) treats as orthonormal. These overlaps vanish for unequal spin parity by symmetry. For equal spin parity, interior eigenfunctions sharing a common Bloch boundary parameter are exactly orthogonal on $r\le a$; the pole-by-pole boundary choice of Sec.~\ref{sec:iav-recap} departs from a common parameter by the level-shift differences, and first-order perturbation theory in the Bloch-operator difference gives the order-of-magnitude estimate $S_{nm}\sim\sqrt{\gamma_n^2\gamma_m^2}\,|S_c'|$, built from the same shift-function derivative as the pole residue $Z_n$ of Sec.~\ref{sec:direct}. This parameter is \emph{not} controlled by the width-to-spacing ratio of conditions 1 and 3: $\Gamma_c^n=2P_c\gamma_{nc}^2$ can be small through the penetrability alone while $\gamma_{nc}^2|S_c'|$ remains finite. The diagonal use of the pole basis therefore carries condition 4, the explicit normalization diagnostic
\begin{equation}
Z_n\simeq 1,\qquad \max_{n\neq m}|S_{nm}|\ll 1,
\label{eq:norm-diagnostic}
\end{equation}
evaluated from tabulated reduced widths and Coulomb shift functions in the same pass as conditions 1 and 3.

Condition 2 is the requirement that $W_x$ have negligible matrix elements between the resonance form factors $\{\phi_n\}$ and the smooth nonresonant $x+A$ continuum. The continuum contribution to the absorptive strength is diagnosed, in a specified continuum model, by the dimensionless ratio
\begin{equation}
\epsilon_{\rm cont}\;\equiv\;\frac{\sigma_R^\text{cont}(\Delta E)}{\sigma_R^\text{res}(\Delta E)},
\label{eq:epsilon-cont}
\end{equation}
where $\sigma_R^\text{res}(\Delta E)$ is the contribution to the $x+A$ reaction cross section integrated over the energy window $\Delta E$ from the resonance-pole sector $\{\phi_n\}$, and $\sigma_R^\text{cont}(\Delta E)$ is the nonresonant contribution over the same window (an $R$-matrix background-pole term, a direct nonresonant capture amplitude, or a smooth absorptive optical-model continuum). By construction, $\sigma_R$ is the reaction (nonelastic) cross section carried by $W_x$ in the kernel of Eq.~(\ref{eq:iav-inclusive}), not the three-body inclusive yield; the elastic $x+A$ branch of $\psi_x$ enters $\sigma_\text{elastic}$ rather than $\sigma_R$, so $R$-matrix hard-sphere phase shifts and the real elastic part of any smooth $U_x$ do not contribute to $\sigma_R^\text{cont}$. Within this restriction $\sigma_R^\text{res}+\sigma_R^\text{cont}\simeq\sigma_R^\text{tot}$ over the window, up to the condition-2-bounded pole-continuum interference. The physical source of its smallness at sub-Coulomb energies in light nuclei is the sparse level density around the reference resonance; explicit closure at a specific benchmark is deferred to future work.

Condition 3 is the requirement that the resonance half-width is small against the spacing to the nearest neighbor of any spin and parity, so that each resonance is resolvable as a separate pole of the spectral sum.

%=====================================================
\section{Three-layer width identification}
\label{app:width}
%=====================================================

This appendix derives Layers A and B of the width relations, which together establish $\xi_n^\text{IAV}=\Gamma_n^\text{non-el}/2$ [Eq.~(\ref{eq:width-dict})], and gives the sub-Coulomb identification with the experimental total width.

\emph{Layer A: Pole of the IAV Green's function.} The participant wave function $\psi_x$ of Eq.~(\ref{eq:iav-inclusive}) is the solution of an inhomogeneous Schr\"odinger equation with complex $U_x$ and outgoing-wave boundary conditions, with the source set by the postform interaction~\cite{IAV1985,LeiMoro2015,LeiMoro2018}. The IAV Green's function $G_x(E_x)=(E_x-T_x-U_x)^{-1}$ that resolves $\psi_x$ on the pole basis is therefore a retarded propagator built from the full complex $U_x$, not a projection onto a subspace orthogonal to the elastic $x+A$ channel. The pole self-energy $\Sigma_n^\text{IAV}(E)$ on the form factor $\phi_n$ has its real part absorbed into the experimental pole position $E_n$ by the construction of $\phi_n$ as the interior eigenfunction of $T_x+V_x^R$ at eigenvalue $E_n$ (in the $R$-matrix convention). Its imaginary part carries two contributions: the elastic escape half-width $\Gamma_x^{n}/2$ (equal to $\Gamma_x^{\,\text{sp}}/2$ in the $C^2S=1$ single-particle convention adopted here), generated by the outgoing-wave coupling of $\phi_n$ to the open elastic $x+A$ channel through the real potential alone and present even for $W_x=0$ at $E_n>0$, and the absorptive displacement $\xi_n^\text{IAV}=|\langle\phi_n|W_x|\phi_n\rangle|$ from the diagonal action of $W_x$. Within the diagonal isolated-pole ansatz~(\ref{eq:wx-ansatz}), $G_x$ near the pole therefore takes the form
\begin{equation}
G_x(E_x) \;\simeq\; \frac{|\phi_n\rangle\langle\phi_n|}{E_x - E_n + i\,\Gamma_n^\text{tot}/2},
\label{eq:Gx-pole}
\end{equation}
with $\Gamma_n^\text{tot}/2=\Gamma_x^{n}/2+\xi_n^\text{IAV}$, the pole in the lower complex half-plane (retarded convention). The absorptive route thus carries the one-level total-width profile of Eqs.~(\ref{eq:omega-n}) and~(\ref{eq:direct-spectral}) directly, $W_x$ supplying only the nonelastic numerator of the inclusive yield; the quasibound form with $\xi_n^\text{IAV}$ alone in the denominator, the treatment of Ref.~\cite{LeiMoro2018}, corresponds to neglecting the elastic escape in the pole displacement and is recovered in the sub-Coulomb limit $\Gamma_x^n\to 0$. In practice, $\xi_n^\text{IAV}$ is not evaluated as a direct radial integral on $W_x(r)$, which would require a profile smooth on the resonance scale that the sub-Coulomb regime lacks, but is read from the tabulated partial widths through the Lane-Thomas identification of Layer B below [Eq.~(\ref{eq:Wx-nonelastic})]. Layer A fixes the physical meaning of $\xi_n^\text{IAV}$; Layer B supplies the operational shortcut.

\emph{Layer B: Channel projection of the absorptive operator $W_x$ as a working identification.} The absorptive operator $W_x$ represents flux removal from the elastic $x+A$ channel into its complement, that is, into the open nonelastic channels $c\ne x$. The standard Lane-Thomas $R$-matrix construction~\cite{LaneThomas1958,Tribble2014} identifies the imaginary part of the participant-target level matrix on the pole $\lambda$ (the absorption of flux into open decay channels) with the channel sum $\sum_c P_c\,\gamma_{\lambda c}^2=\tfrac{1}{2}\sum_c\Gamma_\lambda^c$, in which $\Gamma_\lambda^c\equiv 2\,P_c\,\gamma_{\lambda c}^2$ are the FWHM partial widths built from the reduced widths $\gamma_{\lambda c}^2$ and the channel penetration factors $P_c$. In the IAV setting where $W_x$ encodes the flux removed from the elastic channel into the nonelastic decay channels, the corresponding identification at the diagonal pole matrix element is
\begin{equation}
|\langle\phi_n|W_x|\phi_n\rangle| \;=\; \tfrac{1}{2}\sum_{c\ne x}\Gamma_c^n \;=\; \tfrac{1}{2}\,\Gamma_n^\text{non-el},
\label{eq:Wx-nonelastic}
\end{equation}
in which $\Gamma_n^\text{non-el}=\sum_{c\ne x}\Gamma_c^n$ (for the \nuc{19}{F}$\,+\,p$ resonances, $x=p$ and $\Gamma_n^\text{non-el}=\Gamma_\alpha^n+\Gamma_\gamma^n$). The restriction $c\ne x$ in Eq.~(\ref{eq:Wx-nonelastic}) is not a truncation of the Lane-Thomas channel sum but the Feshbach projection that defines the IAV inclusive yield: the elastic $x+A$ channel is retained explicitly, as the asymptotic outgoing-wave part of $\psi_x$ that carries the elastic-breakup amplitude, and $W_x$ acts only on its complement, the nonelastic decay channels $c\ne x$. The level-matrix imaginary part that $W_x$ reproduces is therefore built from the reaction channels alone.

The partial widths $\Gamma_c^n$ on the right-hand side are the experimentally measured (or $R$-matrix-fitted) decay-channel widths of the resonance $n$, used directly under the observed-parameter convention. Layers A and B give the same $\xi_n^\text{IAV}=\Gamma_n^\text{non-el}/2$ because each computes the rate at which $W_x$ removes the resonance from the elastic channel into the nonelastic decay channels: for a state prepared as $\psi(0)=\phi_n$ and evolving under the complex participant-target potential, the pole occupation decays as $\mathrm{d}|\langle\phi_n|\psi\rangle|^2/\mathrm{d}t=(2/\hbar)\langle\phi_n|W_x|\phi_n\rangle\,|\langle\phi_n|\psi\rangle|^2$, so the instantaneous rate at $t=0$ is $-\Gamma_n^\text{non-el}/\hbar$ (Layer A), while the Lane-Thomas penetrability sum over the same nonelastic channels gives the identical value (Layer B).

\emph{Sub-Coulomb identification with the experimental total width.} The absorptive half-width $\xi_n^\text{IAV}=\Gamma_n^\text{non-el}/2$ is a distinct object from the experimental total width $\Gamma_n^\text{tot}=\Gamma_x^n+\Gamma_n^\text{non-el}$: the former is the absorption out of the elastic channel, the latter includes the elastic re-emission $\Gamma_x^n$. The two halves coincide whenever $\Gamma_x^n\ll\Gamma_n^\text{non-el}$, the generic sub-Coulomb case for charged participants because $\Gamma_x^n$ carries the participant-channel Coulomb penetration factor $\sim\exp(-2\pi\eta_{xA})$ at large Sommerfeld parameter $\eta_{xA}\gg 1$; the identification~(\ref{eq:width-dict}) is the strict equality supplemented by this approximation, controlled by the ratio $\Gamma_x^n/\Gamma_n^\text{tot}$.

Specializing to the \nuc{19}{F}$\,+\,p$ application (so that $x=p$ and $\Gamma_x^n\equiv\Gamma_p$), at the $11$~keV $1^+$ resonance the proton partial width $\Gamma_p\approx 1.1\times 10^{-28}~\text{eV}$~\cite{deBoer2021} and the dominant $\alpha_2$ partial width $|\Gamma_{\alpha_2}|\approx 590~\text{eV}$~\cite{ZhangJUNA2022Nature}, together with the upper bound $\Gamma_{\gamma_1}<5~\text{eV}$~\cite{ZhangJUNA2022Nature} on the $\gamma$ channel, give $\Gamma_n^\text{tot}\simeq |\Gamma_{\alpha_2}|\sim 6\times 10^2~\text{eV}$ and $\Gamma_p/\Gamma_n^\text{tot}\approx 2\times 10^{-31}$, so the sub-Coulomb identification holds with negligible correction. At the higher reference resonances of the literature two-resonance normalization the suppression is weaker, $\Gamma_p/\Gamma_n^\text{tot}\approx 1.6\%$ at $323$~keV~\cite{deBoer2021,Su2025} and weaker still at $828$~keV; there the quasibound substitution $\xi_n^\text{IAV}\simeq\Gamma_n^\text{tot}/2$, if used, introduces a correction comparable to the dynamical-approximation content of Sec.~\ref{sec:postform} and must then be propagated in the same systematic budget; the full profile of Eq.~(\ref{eq:omega-n}) carries no such omission.

\bibliography{refs}

\end{document}